\RequirePackage[2020-02-02]{latexrelease}
\documentclass[reprint,superscriptaddress,amsmath,amssymb,aps,prc,twocolumn,longbibliography,%groupedaddress,%unsortedaddress,%runinaddress,%frontmatterverbose,%preprint,%preprintnumbers,%nofootinbib,%nobibnotes,%bibnotes,%prb,%rmp,%prstab,%prstper,%floatfix
]{revtex4}

\usepackage{graphicx,color}% Include figure files
\usepackage{dcolumn}% Align table columns on decimal point
\usepackage{bm}% bold math
\usepackage{hyperref}% add hypertext capabilities
\usepackage[dvipsnames]{xcolor}
\usepackage{amsmath}

 \newcommand {\nc} {\newcommand}
 \nc {\Sec} [1] {Sec.~\ref{#1}}
 \nc {\IR} [1] {\textcolor{red}{#1}} 
 \nc {\IB} [1] {\textcolor{blue}{#1}}
 \nc {\IV} [1] {\textcolor{violet}{#1}} 
 \nc {\IG} [1] {\textcolor{green}{#1}}
\begin{document}

\title{Prediction of (p,n) Charge-Exchange Reactions with Uncertainty Quantification}

\author{T. R. Whitehead}
\affiliation{Facility for Rare Isotope Beams, Michigan State University, East Lansing, Michigan 48824, USA}
\author{T. Poxon-Pearson}
\affiliation{Facility for Rare Isotope Beams, Michigan State University, East Lansing, Michigan 48824, USA}
\affiliation{Department of Physics and Astronomy, Michigan State University, East Lansing, Michigan 48824, USA}
\author{F. M. Nunes}
\affiliation{Facility for Rare Isotope Beams, Michigan State University, East Lansing, Michigan 48824, USA}
\affiliation{Department of Physics and Astronomy, Michigan State University, East Lansing, Michigan 48824, USA}
\author{G. Potel}%
\affiliation{Facility for Rare Isotope Beams, Michigan State University, East Lansing, Michigan 48824, USA}
\affiliation{Lawrence Livermore National Laboratory L-414, Livermore, California 94551, USA}

\date{\today}

\begin{abstract}
\begin{description}
\item[Background]
Charge-exchange reactions are a powerful tool for exploring nuclear structure and nuclear astrophysics, however, a robust charge-exchange reaction theory with quantified uncertainties is essential to extracting reliable physics. 
\item[Purpose] The goal of this work is to determine the uncertainties due to optical potentials used in the theory for charge-exchange reactions to isobaric analogue states.
\item[Method] We implement a two-body reaction model to study (p,n) charge-exchange transitions and perform a Bayesian analysis. The (p,n) reaction to the isobaric analog states of $^{14}$C, $^{48}$Ca, and $^{90}$Zr targets are studied over a range of beam energies. We compare predictions using standard phenomenological optical potentials with those obtained microscopically.
\item[Results] Charge-exchange cross sections are reasonably reproduced by modern optical potentials. However, when uncertainties in the optical potentials are accounted for, the resulting predictions of charge-exchange cross sections have very large uncertainties.
\item[Conclusions] The charge-exchange reaction cross section is strongly sensitive to the input interactions, making it a good candidate to further constrain nuclear forces and aspects of bulk nuclear matter. However, further constraints on the optical potentials are necessary for a robust connection between this tool and the underlying isovector properties of nuclei.
\end{description}
\end{abstract}


\maketitle

\begin{figure*}
\begin{center}
\includegraphics[scale=0.25]{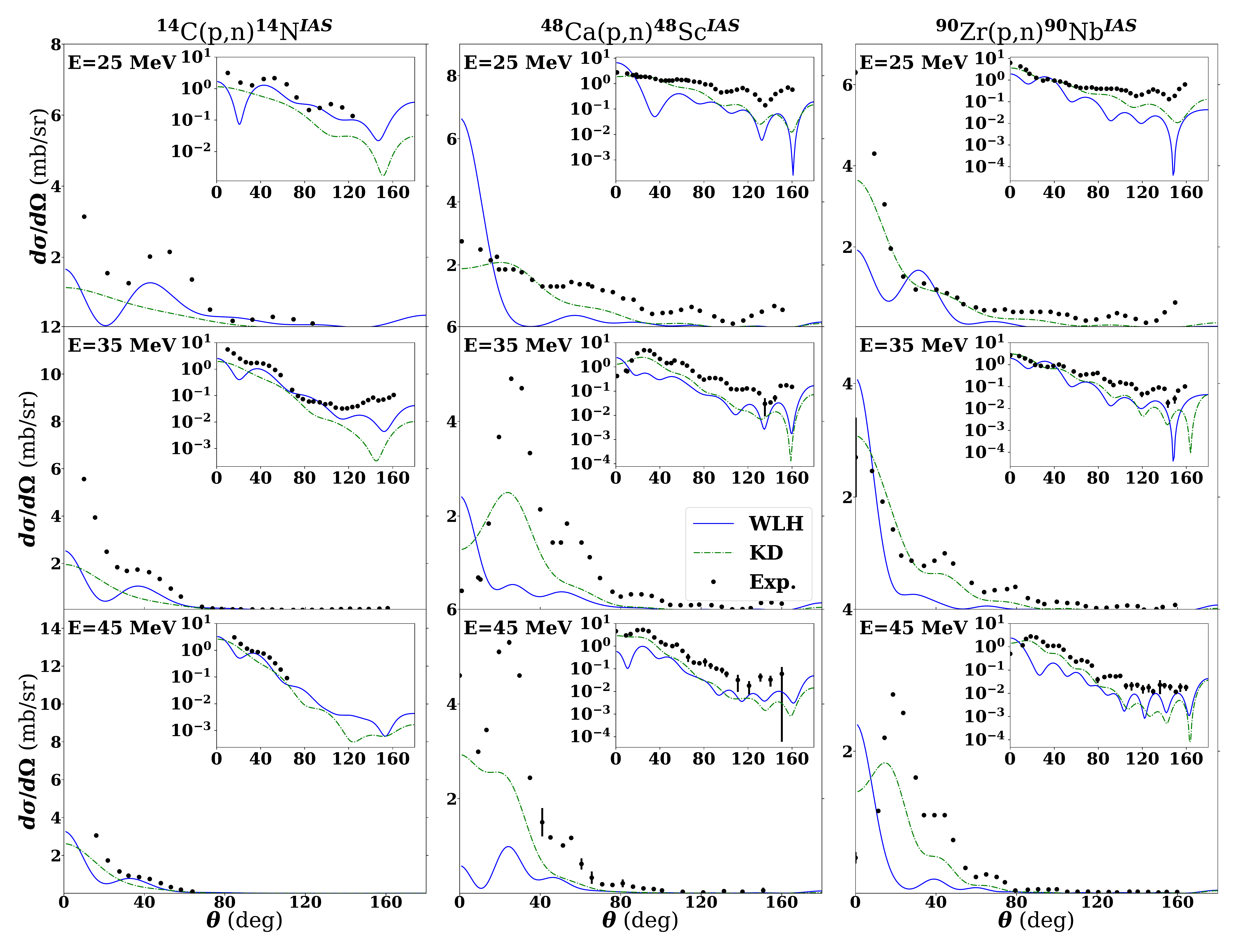}
\caption{Angular distributions for $^{14}$C(p,n)$^{14}$N$^{IAS}$, $^{48}$Ca(p,n)$^{48}$Sc$^{IAS}$, and $^{90}$Zr(p,n)$^{90}$Nb$^{IAS}$ at $E = 25, 35, 45$ MeV. Insets show the same results in log scale. The curves represent predictions from the WLH (blue solid) and KD (green dash-dot) global optical potentials. Experimental data with error bars (black) are from \cite{Taddeucci1984,Doering1975}.
\label{fig:CHEX1}}
\end{center}
\end{figure*}

\section{\label{sec:level1}Introduction}
Charge-exchange reactions are isobaric transitions where a neutron in the target is exchanged with a proton in the projectile, or vice-versa. Charge-exchange reactions are a versatile probe with broad applications ranging from testing nuclear structure calculations to constraining neutrino detector response \cite{Fujiwara2000, Ejiri1998}. Since they can populate the same states as beta decay, these reactions can be used to extract information about weak decays in regions where beta decay cannot be directly observed. Moreover, charge-exchange reactions are a powerful tool in the realm of nuclear astrophysics: they can provide constraints on bulk properties of nuclear matter, which are central to understanding neutron stars and their mergers, and act as an indirect probe for stellar electron-capture processes.

Charge-exchange reactions can be used to constrain bulk nuclear matter by placing limits on the nuclear symmetry energy \cite{Krasznohorkay1999}. The symmetry energy accounts for the energy difference due to an imbalance of neutrons and protons within nuclear matter and is directly linked to the nuclear equation of state, a key component for modeling the behavior of neutron stars. Fermi transitions ($\Delta$L=0, $\Delta$S=0, $\Delta$T=1) between IAS provide a unique tool for exploring isovector densities. A significant number of experimental groups have measured charge-exchange cross sections to IAS for a wide range of isotopes (see for example \cite{carlson1975,ohnuma1986,ohnuma1987,Jon1997,Jon2000}).
Moreover, there have been several theoretical efforts, informed by these measurements, to explore the isovector properties of nuclei \cite{Danielewicz2017,Khoa2014, Loc2017, Loc2014}.

Typically charge-exchange reactions to the IAS are studied within 1-step DWBA. Models that incorporate more complex reaction dynamics explored in other reaction channels, such as transfer reactions and Coulomb dissociation (see \cite{capel2017, Titus20162, Li2018, Deltuva2005, Hlophe2019}), could present relevant extensions for improving the theory of charge-exchange reactions to IAS. More importantly, given the advances in uncertainty quantification for reactions \cite{Lovell2018,king2019,catacora2019,lovell2020}, it is imperative to quantify the uncertainties in the charge-exchange predictions from theory. To achieve this, uncertainties of the optical potential, the principal input for predicting charge-exchange reactions, must be understood. 

Over the last decade much work has been invested in obtaining microscopic optical potentials that are derived from many-body calculations based on realistic nuclear forces (e.g \cite{rotureau2017,rotureau2018,idini2019,burrows20,whitehead21}). In particular, the Whitehead-Lim-Holt (WLH) optical potential \cite{whitehead21}, derived from many-body perturbation theory with chiral forces in nuclear matter, is the only microscopic {\it global} optical potential. It has a range in mass and energy comparable to the most recent phenomenological global parameterization of Koning and Delaroche (KD) \cite{Koning2003}. When compared to experimental data for elastic scattering, the WLH potential and the KD potential have similar performance \cite{Whitehead19,Whitehead20}.

Ultimately, our goal is to build a modern framework for charge-exchange reactions from scratch. As a first step, we begin with Fermi transitions within a one-step reaction model. We examine the sensitivity of the charge-exchange cross section to the interaction that mediates the process and compare results obtained using a phenomenological optical potential and the global microscopic WLH potential. For the phenomenological approach, we use the Bayesian tools developed in \cite{Lovell2018} and propagate the uncertainties to the charge-exchange cross section. For the microscopic approach, we use the optical potential derived in \cite{whitehead21} to propagate the uncertainties from chiral forces to the charge-exchange cross section. 
The paper is organized in the following way. In Section \ref{sec:theory} we briefly describe the necessary theory and numerical details. The charge-exchange results with uncertainty quantification are presented in Section \ref{sec:results} and the uncertainty quantification is described in Section \ref{sec:mcmc}. A discussion of uncertainties in the prediction of charge-exchange reactions is given in Section \ref{sec:discussion}. We draw conclusions of this investigation in Section \ref{sec:conclusions}.

\begin{figure*}
\begin{center}
\includegraphics[scale=0.17]{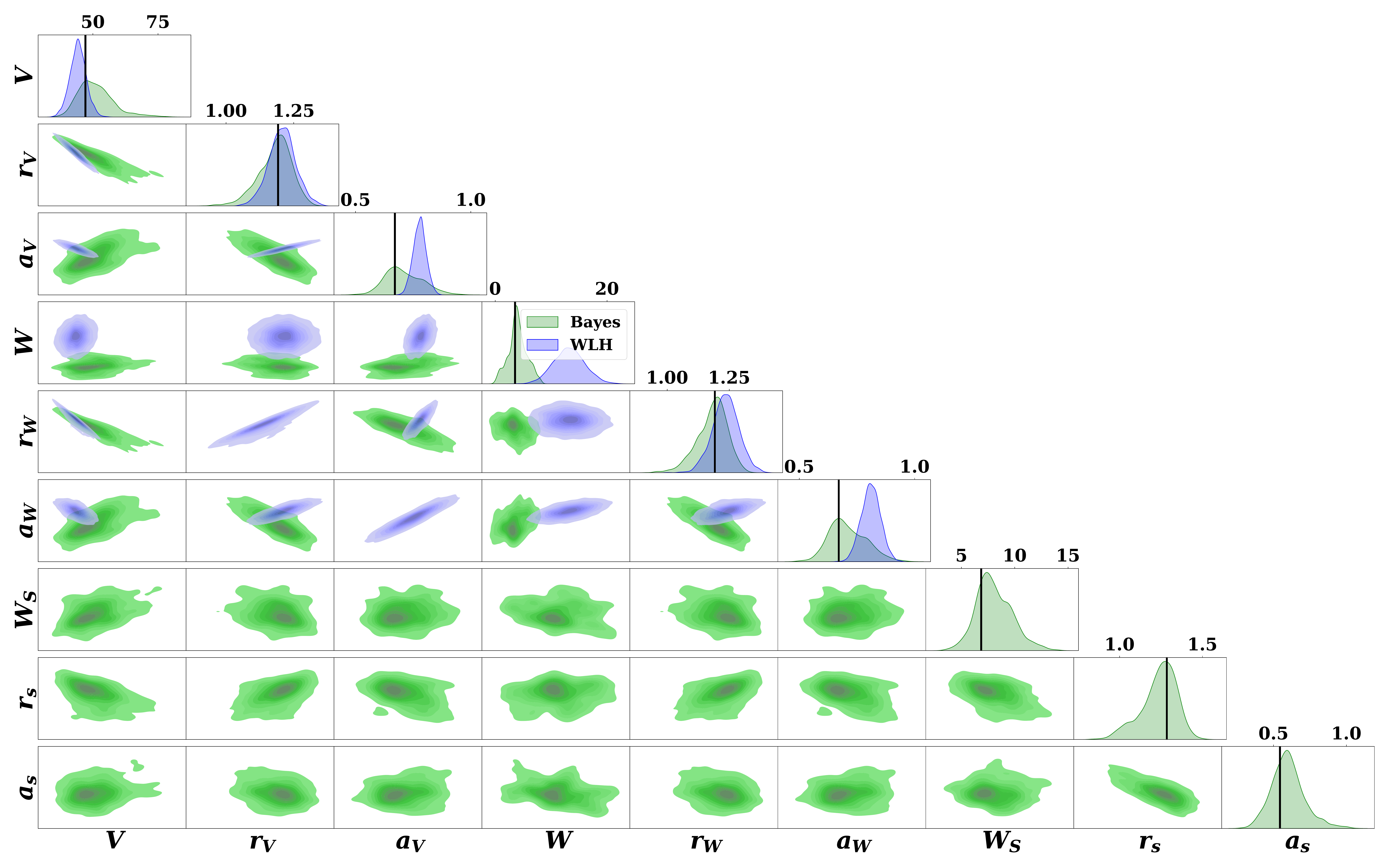}
\caption{Histogram and correlation plots of optical potential parameters for p+$^{48}$Ca at $E = 35$ MeV that characterize the entrance channel for the $^{48}$Ca(p,n)$^{48}$Sc$^{IAS}$ reaction at a beam energy of $E = 35$ MeV. Parameter distributions of the Bayesian posteriors are shown in green and distributions of the WLH potential are shown in blue. The black vertical lines indicate the mean value of the Bayesian prior which is taken from the Koning-Delaroche global optical potential.
\label{fig:param1}}
\end{center}
\end{figure*}

\begin{figure*}
\begin{center}
\includegraphics[scale=0.17]{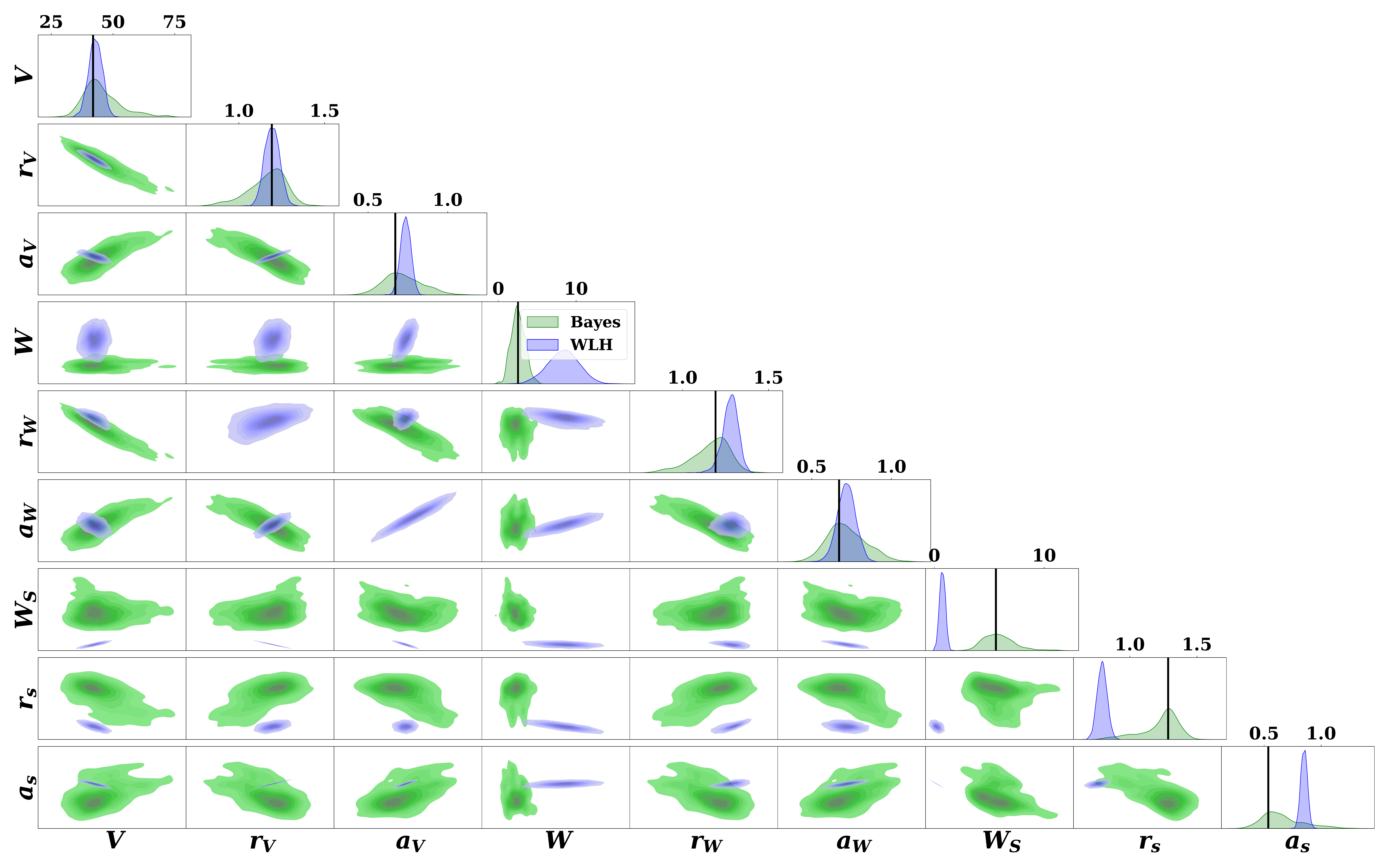}
\caption{Histogram and correlation plots of optical potential parameters for n+$^{48}$Sc at $E = 28$ MeV that characterize the exit channel for the $^{48}$Ca(p,n)$^{48}$Sc$^{IAS}$ reaction at a beam energy of $E = 35$ MeV. Parameter distributions of the Bayesian posteriors are shown in green and distributions of the WLH potential are shown in blue. The black vertical lines indicate the mean value of the Bayesian prior which is taken from the Koning-Delaroche global optical potential.
\label{fig:param2}}
\end{center}
\end{figure*}

%%%%%%%%%%%%%%%%%%%%%%%%%%%%%%%%%%%%%%%%%%%%%%%%%%%%%%%%%
\section{\label{sec:theory}Theoretical Framework}

Within this work we have employed a two-body framework using single-step DWBA, to analyze (p,n) charge-exchange reactions between $0^+$ IAS, as done in previous studies (e.g. \cite{Danielewicz2017}). In this formalism, charge-exchange transitions are mediated by the difference between the incoming and outgoing optical potentials which characterize nucleon-target elastic scattering. These optical potentials may be written in the Lane form \cite{Lane1962}:

\begin{equation} \label{Lane}
U(R_{1A})=U_0(R_{1A}) +\frac{\bm{\tau \cdot T}}{4A}U_1(R_{1A}),
\end{equation}
where $R_{1A}$ is the coordinate that connects the projectile to the center of mass of the target, $U_0(R_{1A})$ is the isoscalar potential which drives elastic scattering, $U_1(R_{1A})$ is the isovector potential which drives the charge-exchange transition, and $\bm{\tau}$ and $\bm{T}$ are the isospin operators which act on the projectile nucleon and target nucleus, respectively. The isovector part of the Lane potential can be written as

\begin{equation}
U_1(R_{1A})=\frac{A}{2(N-Z)} [U_n(R_{1A})-U_p(R_{1A})],
\label{eq-lane}
\end{equation}
where N, Z, and A are the neutron number, charge number, and mass number of the target nucleus and $U_n$, $U_p$ are the neutron- and proton-target optical potentials. Here we will utilize the global potentials developed by Koning-Delaroche (KD) \cite{Koning2003} and WLH \cite{whitehead21}. Both potentials take the following form

\begin{align}
	U_{n,p}(R) &= -V f(R;r_V,a_V) - i W f(R;r_W,a_W) \\ \nonumber 
	&+i 4a_S W_S \frac{d}{dR} f(R;r_S,a_S) \\ \nonumber 
	&+V_{SO} \frac{1}{m_\pi^2}\frac{1}{R} \frac{d}{dR} f(R;r_{SO},a_{SO}) \vec \ell \cdot \vec \sigma,
\end{align}
where $f(R;r_i,a_i)=\frac{1}{1+e^{(R-A^{1/3}r_i)/a_i}}$ (for $V$ and $W$) and its derivative (for $W_S$ and $V_{SO}$). Note that the KD potential also includes an imaginary spin-orbit term that is negligible for both elastic and charge-exchange channels at the scattering energies presently considered. This term is included for all calculations labeled as Koning-Delaroche or Bayesian.

\begin{figure*}
\begin{center}
\includegraphics[scale=0.22]{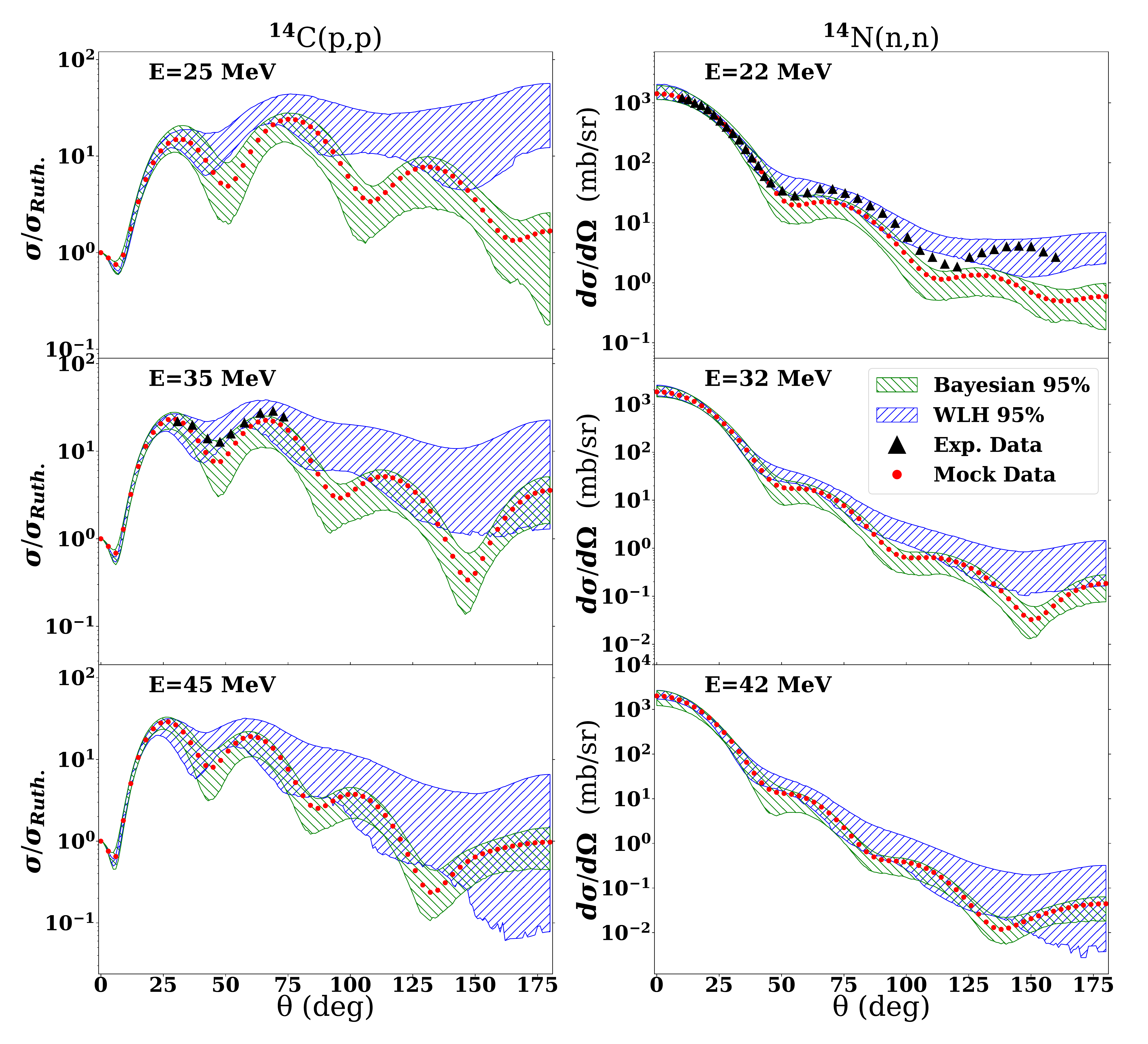}
\caption{Angular distributions for $^{14}$C(p,p) at $E = 25, 35, 45$ MeV and $^{14}$N(n,n) at $E = 22, 32, 42$ MeV. The Bayesian 95\% confidence interval is shown in green and the 95\% confidence interval of the WLH optical potential is shown in blue. Experimental data are from \cite{YASUE90,Olsson89}. Mock data are predictions of the Koning-Delaroche global optical potential.
\label{fig:elastic1}}
\end{center}
\end{figure*}

\begin{figure*}
\begin{center}
\includegraphics[scale=0.22]{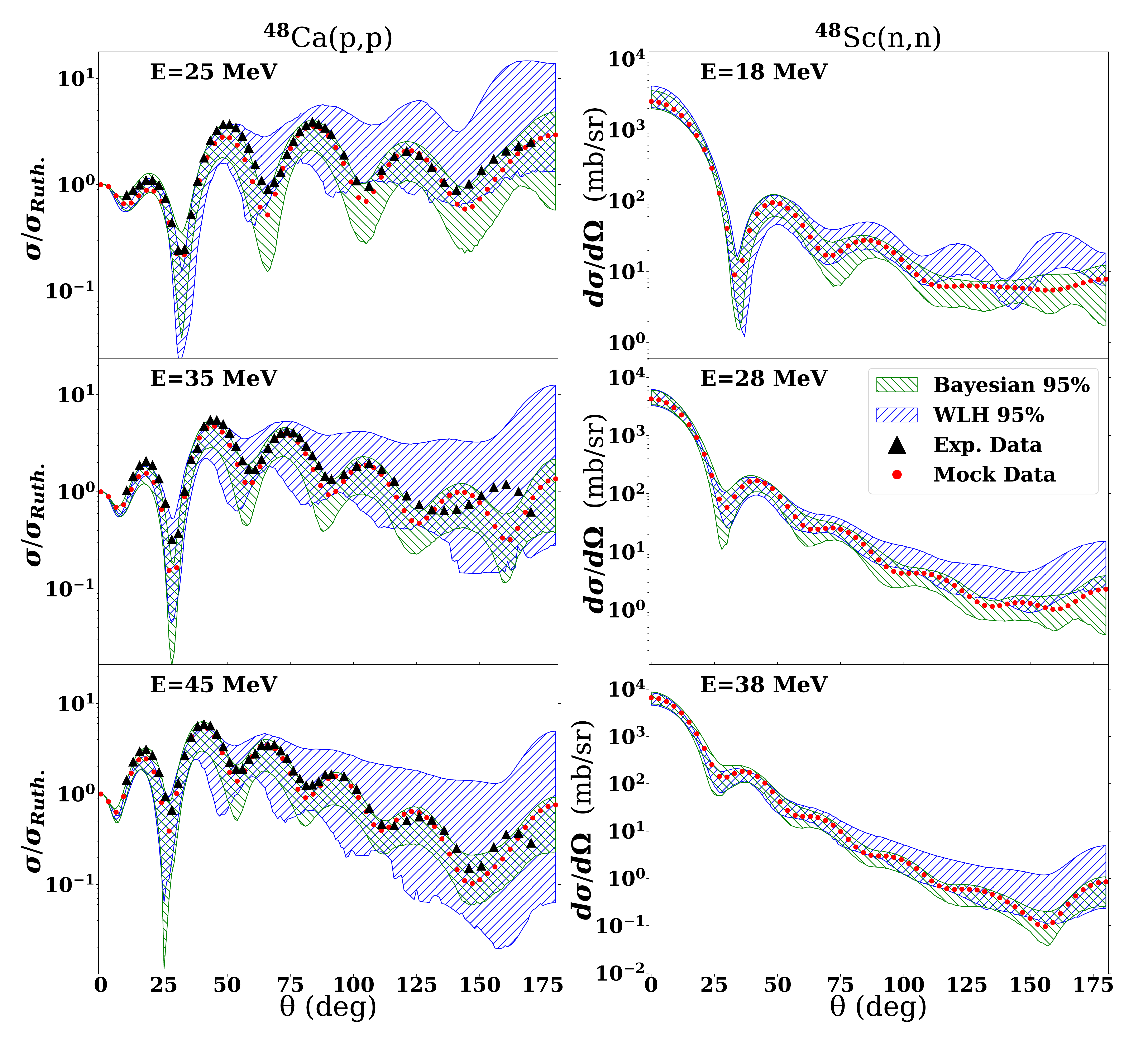}
\caption{Angular distributions for $^{48}$Ca(p,p) at $E = 25, 35, 45$ MeV and $^{48}$Sc(n,n) at $E = 18, 28, 38$ MeV. The Bayesian 95\% confidence interval is shown in green and the 95\% confidence interval of the WLH optical potential is shown in blue. Experimental data are from \cite{McCamis86}. Mock data are predictions of the Koning-Delaroche global optical potential.
\label{fig:elastic2}}
\end{center}
\end{figure*}

\begin{figure*}
\begin{center}
\includegraphics[scale=0.22]{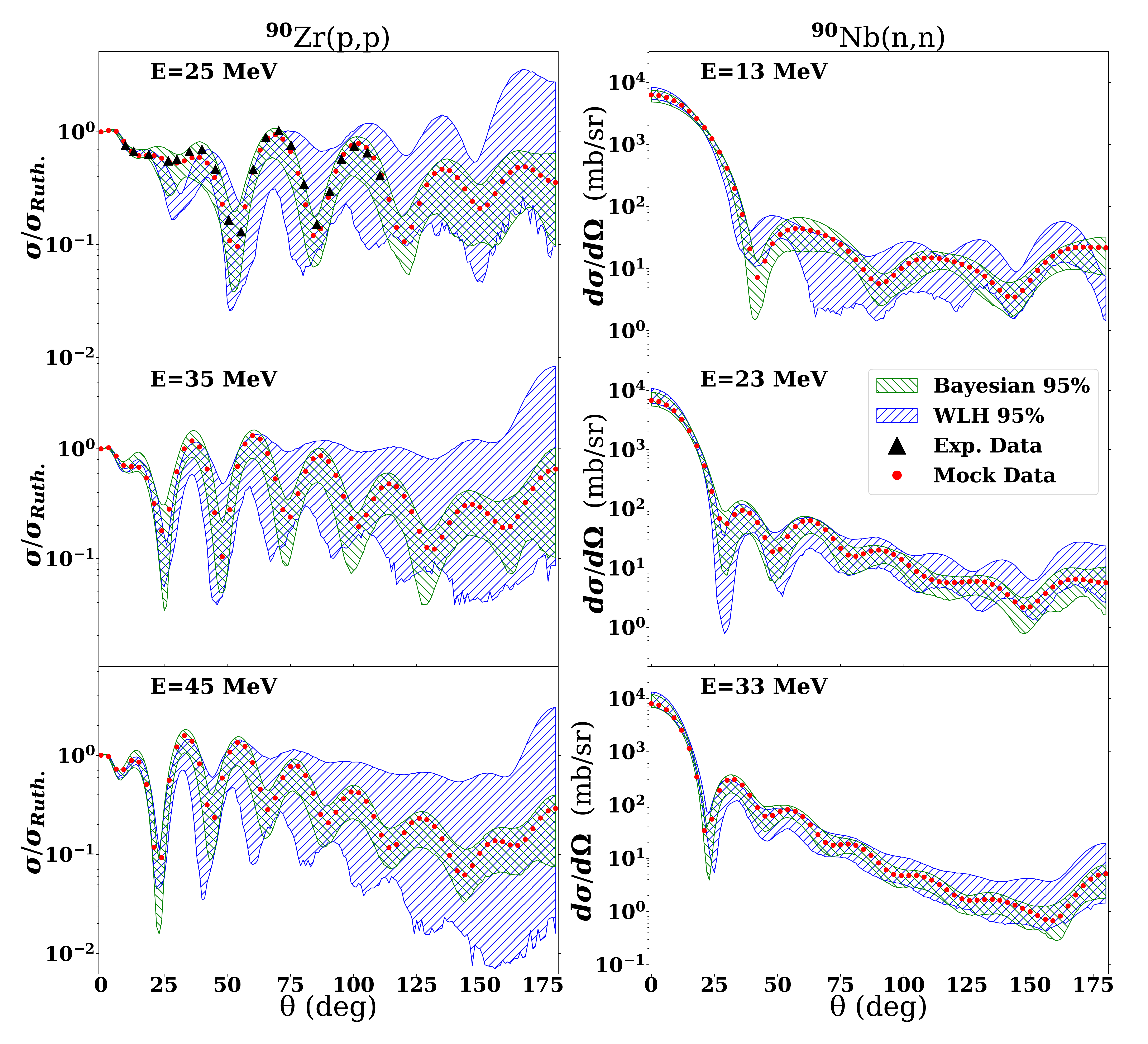}
\caption{Angular distributions for $^{90}$Zr(p,p) at $E = 25, 35, 45$ MeV and $^{90}$Nb(n,n) at $E = 13, 23, 33$ MeV. The Bayesian 95\% confidence interval is shown in green and the 95\% confidence interval of the WLH optical potential is shown in blue. Experimental data are from \cite{VANDERBIJL83}. Mock data are predictions of the Koning-Delaroche global optical potential.
\label{fig:elastic3}}
\end{center}
\end{figure*}

In the case of (p,n) charge-exchange reactions to the isobaric analog state, the transition matrix element can be expressed simply as
\begin{equation}
T^{CE}_{2B}=\langle \chi_f(R_{1A}) | 2\frac{\sqrt{|N-Z|}}{A}U_1(R_{1A}) | \chi_i(R_{1A}) \rangle,
\label{tmatrix}
\end{equation}
where $\chi_i(R_{1A})$ and $\chi_f(R_{1A})$ 
represent the p+A/n+B distorted waves that are calculated using $U_p$ and $U_n$, respectively. Note that this simplified matrix element is specific to (p,n) IAS transitions. A more complex formulation is needed for other charge-exchange reactions (see Ref. \cite{Taddeucci1987})

Calculations for (p,n) reactions are performed using a new charge-exchange reaction code, {\sc chexpn}, that was developed for this work. This code is not to be confused with {\sc chex2} \cite{CLARKE86}, a charge-exchange code capable of calculating form factors for reactions with composite probes in addition to (p,n). Scattering wave functions $\chi_{i,f}$, as well as elastic scattering cross sections, were benchmarked against results produced by the reaction code {\sc fresco} \cite{FRESCO}. Charge-exchange calculations were compared with results from the unpublished code used in \cite{Danielewicz2017}. For all cases presented here, we began by calculating the charge-exchange cross section using two choices for the optical potential (KD and WLH). For consistency, the same optical potential was used to calculate both the distorted waves and the transition operator in Eq. \ref{tmatrix}. Calculations for each of the cases studied here required less than 10 partial waves to converge and take only a few seconds to run.

%%%%%%%%%%%%%%%%%%%%%%%%%%%%%%%%%%%%%%%%%%%%%%%%%%%%%%%%%
\section{\label{sec:results}Prediction of charge-exchange without uncertainty quantification}

We study (p,n) charge-exchange reactions to $0^+$ IAS in $^{14}$C, $^{48}$Ca, and $^{90}$Zr. These targets were chosen because they span a large range of nuclear masses and there are experimental charge-exchange cross section data available for the IAS transition of each target. Each of these reactions are studied at three different beam energies: $E$=25, 35, and 45 MeV.

Experimental cross sections for $^{14}$C(p,n) are taken from Taddeucci et al., where the reaction to the IAS was measured at $E_{lab}=$ 25.7, 35 and 45 MeV \cite{Taddeucci1984}. Experimental charge-exchange cross sections for $^{48}$Ca(p,n) and $^{90}$Zr(p,n) come from Doering et al., in which transitions were measured at $E_{lab}=$ 25, 35 and 45 MeV for both targets \cite{Doering1975}.

In the first part of this study, we calculate IAS charge-exchange transitions using the global optical potentials referenced in Sec. \ref{sec:theory}. The cross sections for each of the reactions considered in this work are shown in Fig. \ref{fig:CHEX1}: first column is for $^{14}$C(p,n)$^{14}$N$^{IAS}$ , the second column is for $^{48}$Ca(p,n)$^{48}$Sc$^{IAS}$ , and the third column for $^{90}$Zr(p,n)$^{90}$Nb$^{IAS}$. The three rows correspond to the different beam energies, $E_{lab}=$ 25, 35, and 45 MeV. Insets show the same information on a log scale, which is particularly helpful for clarifying behavior at large angles where cross sections are typically small. The solid blue lines correspond to predictions using the microscopic potential WLH while the dashed green lines are predictions of the KD phenomenological global potential. The results of the WLH potential in this section pertain to a particular chiral interaction (that is computed up to N$^{3}$LO with a cutoff of $\Lambda=450$ \cite{Whitehead20}) for direct comparison to the Koning-Delaroche results.

When comparing cross sections produced by the two global optical potentials, it is clear that the choice of optical potential has a large impact on the angular distributions. Despite having similar orders of magnitude, the angular distributions obtained using KD and WLH are very different, in part due to differences in their radii. It is clear from Fig. \ref{fig:CHEX1} that if we were to estimate the error from comparing KD and WLH, it would exceed the standard 30$\%$ uncertainty that is often cited in reactions \cite{Nunes2011}. A more rigorous quantification of the uncertainty is necessary and will be done in the next section.

Next we compare our predictions to experimental data. As can be seen in Fig. \ref{fig:CHEX1}, in most cases the charge-exchange calculations are able to capture the broad features of the data at small and large angles. Although a particular potential may describe any specific data set better than the other, the opposite may be true for a different target/energy combination. Phenomenological optical potentials are fit to large data sets and optimized to best describe trends in elastic scattering data over a wide range of energies and target masses. Even for the elastic channel, they cannot predict all the details in the angular distributions.

We note that our study employs a similar methodology to that employed by Danielewicz et al. \cite{Danielewicz2017}. In that work, an equivalent charge-exchange framework is used to calculate transitions between $0^+$ IAS using a Lane potential. The results in this section are consistent with those presented by Danielewicz et al., however, we do not include any modifications to the isovector term due to Coulomb interactions since Coulomb interactions are fully included in our calculations. See the discussion pertaining to Eq. (24) and (25) in Ref. \cite{Danielewicz2017} for more details.

%%%%%%%%%%%%%%%%%%%%%%%%%%%%%%%%%%%%%%%%%%%%%%%%%%%%%%%%%
\section{\label{sec:mcmc}Uncertainty Quantification}

\begin{figure*}
\begin{center}
\includegraphics[scale=0.25]{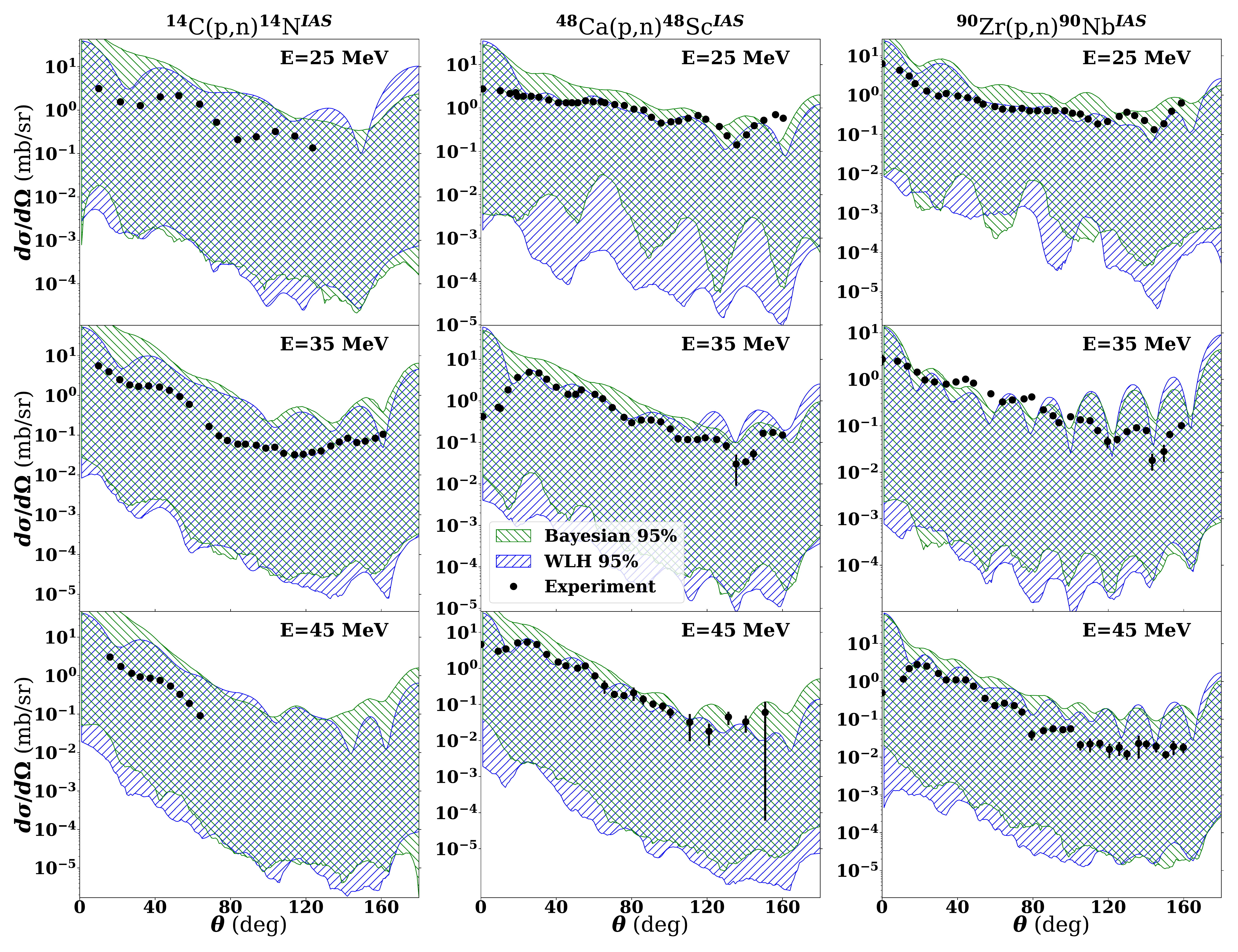}
\caption{Angular distributions for $^{14}$C(p,n)$^{14}$N$^{IAS}$, $^{48}$Ca(p,n)$^{48}$Sc$^{IAS}$, and $^{90}$Zr(p,n)$^{90}$Nb$^{IAS}$ at $E = 25, 35, 45$ MeV. The Bayesian 95\% confidence interval propagated from elastic scattering is shown in green and the 95\% confidence interval of the WLH optical potential is shown in blue. Experimental data with error bars shown in black are from \cite{Taddeucci1984,Doering1975}. Note that the local minima of the charge-exchange cross sections cause the lower band to be jagged; these features have been smoothed out so that the uncertainty bands may be more clearly interpreted.
\label{fig:CHEX2}}
\end{center}
\end{figure*}

We present two distinct approaches of estimating the theoretical uncertainty in predictions of charge-exchange cross sections. Firstly, from the phenomenological perspective, we carry out a full Bayesian analysis by constraining the optical potentials to elastic scattering and propagating the uncertainties to charge-exchange. Secondly, from a microscopic perspective, we utilize the WLH global optical potential that estimates the theoretical uncertainty of scattering observables based on uncertainties from chiral interactions.

In our Bayesian analysis, we use elastic scattering mock data to constrain the optical potential and then, from the obtained parameter distributions, propagate those uncertainties to obtain confidence intervals on the charge-exchange cross sections. For more details on the implementation of the Bayesian analysis, see Ref. \cite{Lovell2018}. The entrance and exit channel interactions are modeled by proton- and neutron-target optical potentials at the appropriate energies. Priors for the optical potential parameters are gaussians centered on the Koning-Delaroche values with widths equal to the mean value of the distribution, reflecting that \textit{a priori} there is a large uncertainty in optical potential parameters. We find that Bayesian optimization of all three central optical potential terms is not adequately constrained by elastic scattering data. We remedy this by assuming that the radius and diffuseness parameters of the real volume and imaginary volume terms are equal, following the assumptions in \cite{Koning2003}. The resulting seven parameter Bayesian optimization yields improved posterior distributions. For each reaction, posteriors are generated by sixteen separate MCMC samplings, each with 1600 parameter pulls, to ensure that parameter space is sufficiently explored. Ideally, proton and neutron elastic scattering data would be available for a wide range of energies and targets, however, there are large gaps in the available data. Because of this, we use mock data generated from the predictions of the Koning-Delaroche global optical potential. It should be noted that although this mock data is expected to be close to empirical data, there are discrepancies in some cases. However, they are of little consequence to the uncertainties obtained \cite{CatacoraRios21}. In contrast to our approach that only uses elastic scattering to constrain the optical potential, Danielewicz et al., used charge-exchange data in combination with proton and neutron elastic scattering.

We now turn to the uncertainty quantification in the microscopic approach. Like traditional global optical potentials, the WLH global optical potential is expressed in terms of Woods-Saxon parameters that define the strength and shape of the various terms in the potential. However, the WLH potential of \cite{whitehead21} is unique since it is the first global optical potential to incorporate uncertainty quantification. The WLH optical potential is based on five separate global optical potentials, each calculated using a different chiral interaction. Each of these five global optical potentials is parameterized in the same way, then for each parameter a normal distribution is assumed about the values they take in each of the five global optical potentials. In practice, one generates uncertainty estimates by randomly sampling the WLH optical potential parameter distributions many times and calculating a given reaction observable for each sample. The resulting calculations represent the uncertainty in that observable, arising from chiral EFT uncertainties through the optical potential. This procedure is carried out for charge-exchange reactions by sampling the WLH potential for both the entrance and exit channel optical potentials and calculating the cross section for each sample. Samples from the WLH potential include twelve Woods-Saxon parameters from the real volume, imaginary volume, imaginary surface, and real spin-orbit terms. The resulting 95\% confidence intervals are directly compared to the Bayesian results obtained from the phenomenological approach.

\begin{figure*}
\begin{center}
\includegraphics[scale=0.24]{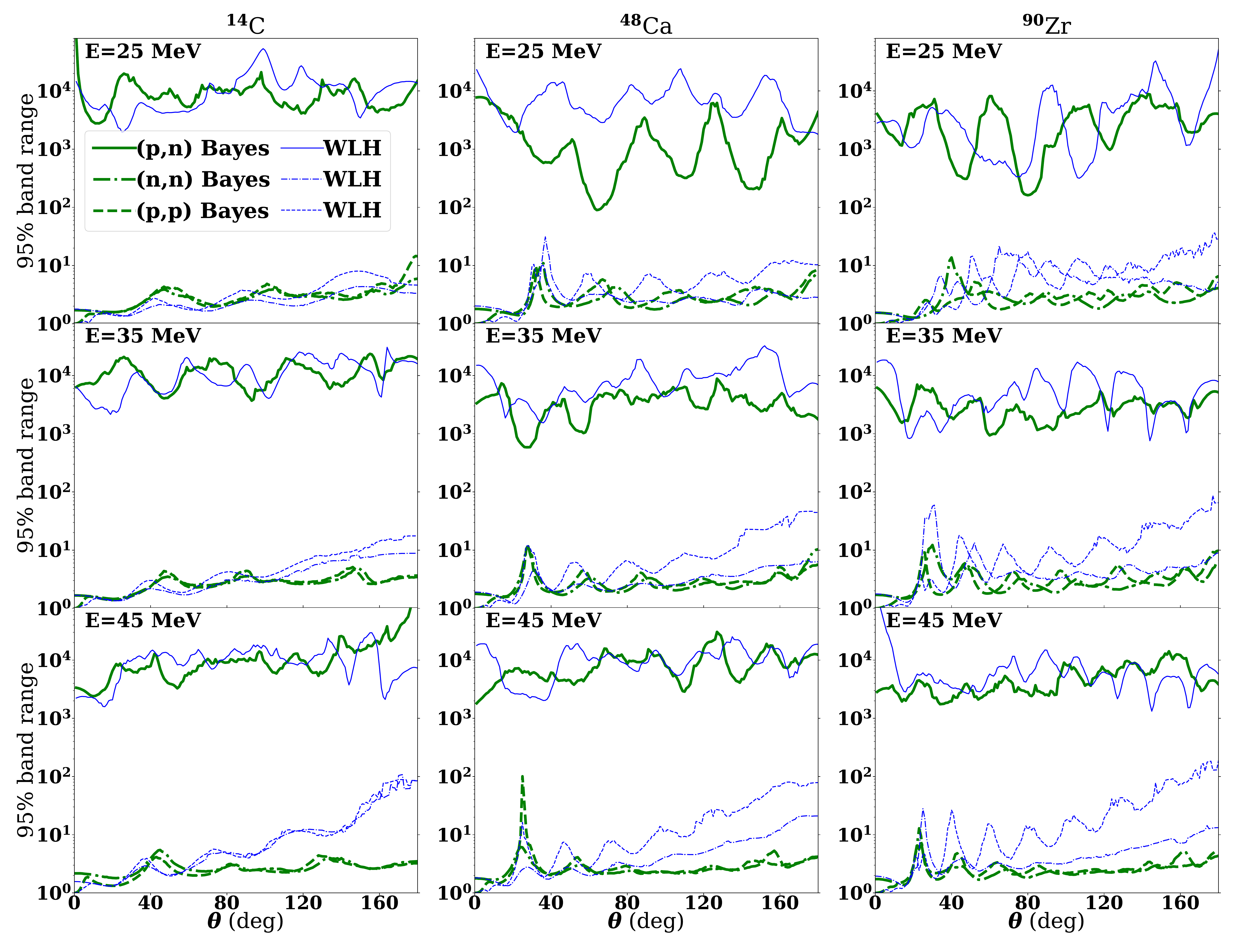}
\caption{Range of the 95\% confidence interval defined as the ratio of the upper limit to the lower limit as a function of the scattering angle. Results for charge-exchange (solid), neutron elastic (dash-dot), and proton elastic (dashed) are shown for the Bayesian (green thick lines) and WLH (blue narrow lines) approaches.
\label{fig:error}}
\end{center}
\end{figure*}

\begin{figure*}
\begin{center}
\includegraphics[scale=0.33]{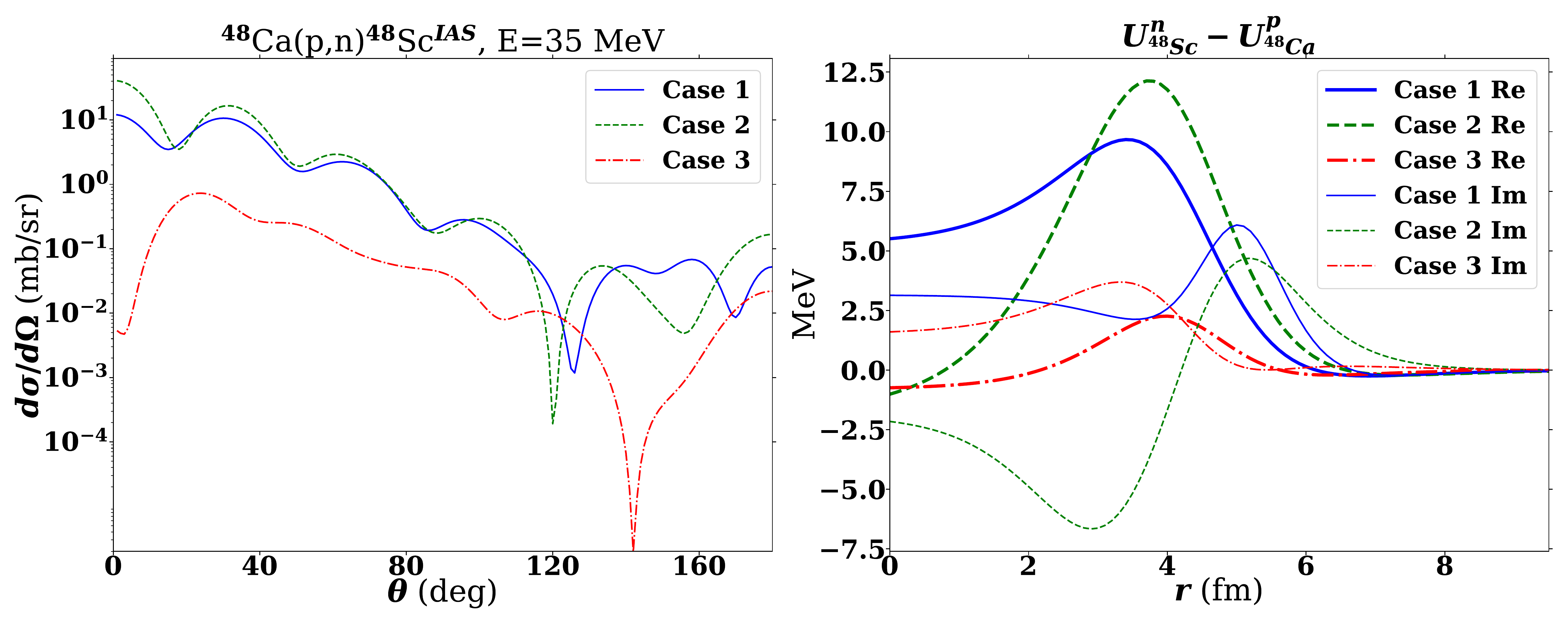}
\caption{Left: Charge-exchange cross sections for three cases taken from the Bayesian posterior for the $^{48}$Ca(p,n)$^{48}$Sc$^{IAS}$ reaction at $E = 35$ MeV. Right: The corresponding real (thick lines) and imaginary (narrow lines) isovector potentials.
\label{fig:comp}}
\end{center}
\end{figure*}
	
In Figs. \ref{fig:param1} and \ref{fig:param2} we show the posterior distributions and correlations for the entrance channel and exit channel optical potentials used in computing the $^{48}$Ca(p,n)$^{48}$Sc$^{IAS}$ cross section for a beam energy of $E = 35$ MeV. This particular reaction is shown as a representative example; other reactions have qualitatively similar features. In Fig. \ref{fig:param1} the real depth parameter distribution from the WLH potential is relatively narrow, while the Bayesian distribution is wider and extends to much higher values. The radius parameter distributions for each approach are similar while the diffuseness parameter distributions in the Bayesian approach are more broad. In the Bayesian approach the real geometry parameters ($r_V,a_V$) are set equal to the imaginary volume geometry parameters ($r_W,a_W$), therefore they do not appear in their correlation plots. Optical potentials derived from nuclear matter calculations generally overestimate the strength of the imaginary term and in this case the WLH imaginary potential is stronger than the Bayesian estimate. The imaginary surface term of the WLH potential diminishes as projectile energy increases and, for protons, it is already zero at $E = 35$ MeV. The Bayesian imaginary surface term is quite strong compared to its imaginary volume and the radius and diffuseness distributions take typical values. The neutron parameter distributions shown in Fig. \ref{fig:param2} follow similar patterns except that the Bayesian radius parameter distributions are wider and the WLH potential has an imaginary surface peak for neutron projectiles at intermediate energies. The WLH imaginary surface is smaller in magnitude than the Bayesian counterpart while the WLH radius distribution is centered on a relatively low value.

In Figs. \ref{fig:elastic1}, \ref{fig:elastic2}, \ref{fig:elastic3} the elastic scattering cross sections corresponding to the entrance and exit channels of the charge-exchange reactions studied in this work are shown. These figures contain the 95\% confidence intervals obtained in the Bayesian approach (green hash), using the mock data from KD (red circles); the 95\% confidence intervals from samples of the WLH potential (blue hash) and, when available, the elastic scattering data (black triangles). The uncertainties on the elastic scattering angular distributions obtained within the Bayesian approach are similar to those presented in previous studies \cite{catacora2019,lovell2020}. The microscopic WLH approach tends to produce similar uncertainties as the phenomenological approach at forward angles, but larger uncertainties at backward angles.

The charge-exchange cross sections with uncertainties from both approaches are shown in Fig. \ref{fig:CHEX2}. In the Bayesian approach, the uncertainties are propagated through optical potentials that are optimized to elastic scattering reactions that represent the entrance and exit channels of the charge-exchange process. In contrast, the uncertainty intervals of the WLH predictions reflect the uncertainties coming from the underlying chiral two- and three-body forces. It is an interesting coincidence that the magnitudes of these uncertainty intervals are so similar, particularly considering that they are derived in a completely different manner. In the following section we discuss the large uncertainties of charge-exchange cross sections.

\section{\label{sec:discussion}Discussion of Charge-Exchange Uncertainty}

The uncertainties for charge-exchange cross sections computed in both frameworks outlined above are surprisingly large. A comparison of the confidence interval widths, taken to be the upper limit of $d \sigma/d\Omega$ divided by the corresponding lower limit, is shown in Fig. \ref{fig:error}: Bayesian results are shown in green while the microscopic WLH results are shown in blue. We include charge-exchange (solid lines), neutron elastic scattering (dash-dot lines) and proton elastic scattering (dashed lines) for all cases considered in this work. We find that the confidence interval widths for charge-exchange in both the Bayesian and WLH approaches are several orders of magnitude larger than the elastic scattering counterparts. 

To understand this unexpected result, we take three example parameter sets from the Bayesian posterior distribution for the $^{48}$Ca(p,n)$^{48}$Sc$^{IAS}$ reaction at $E = 35$ MeV. In Fig. \ref{fig:comp}, we show their charge-exchange predictions as well as the isovector potentials as functions of radius. Case 1 and Case 2 have similar charge-exchange cross sections. Their real and imaginary isovector potentials shown on the right have surface peaks with similar positions and magnitudes, however, their central values are quite different. In contrast, the charge-exchange cross section for Case 3 is substantially different. The real isovector potential for Case 3 has a smaller surface peak than the other cases, but its central value is nearly equal to Case 2. The imaginary isovector potential for Case 3 is shifted towards the interior and slightly smaller than the other cases, while its central values fall between the other cases. 

This examination of the isovector potentials demonstrates that modest differences can lead to very large discrepancies in the charge-exchange cross section. It is important to note that the wide posterior distributions of the Bayesian approach allow for larger differences in the isovector potentials than the three cases presented here. We take examples far from the extremes to show that even minor changes in the surface region of the isovector potential can indeed produce large variations in the charge-exchange cross section. This sensitivity to the isovector potential may be exploited to extract information about the isovector nature of nuclear forces from experimental charge-exchange data.

\section{\label{sec:conclusions}Conclusions}

This study focuses on the predictions for angular distributions of (p,n) charge-exchange reactions to isobaric analogue states. We consider the reaction within a two-body framework, mediated by a Lane potential, corresponding to the difference in the proton- and neutron-target optical potentials. We include two different global optical potentials, one phenomenological \cite{Koning2003} and the other microscopically based \cite{whitehead21}. We study (p,n) reactions on $^{14}$C, $^{48}$Ca, and $^{90}$Zr at $E_{lab}=$25, 35 and 45 MeV. Our calculations show that, although the magnitude obtained with the phenomenological approach is similar to the microscopic approach, the angular distributions differ considerably. At forward angles, the difference greatly exceeds the nominal $30$\% value often quoted.

We also quantify the uncertainties in the charge-exchange cross section. In the phenomenological approach, we perform a Bayesian analysis that incorporates mock elastic scattering data to constrain nucleon-nucleus optical potentials and, from the resulting posterior parameter distributions, obtain the charge-exchange confidence intervals. In the microscopic approach, we utilize the uncertainty quantification integrated in the WLH global optical potential that is based on uncertainties from chiral nuclear forces. We find that the two distinct approaches to quantifying the uncertainty in the charge-exchange angular distributions give similar uncertainty: three to four orders of magnitude. The main conclusion of this work is that the charge-exchange data provides an extraordinarily sensitive constraint to the isovector component of the optical potential. Building on previous work \cite{Danielewicz2017,Love87}, a comprehensive systematic study examining charge-exchange transitions to IAS on a wide range of targets and beam energies has the potential of revealing these details in an remarkable manner.

The large uncertainties found in this work stem at least in part from the construction of the Lane potential from separate proton and neutron interactions. These uncertainties can be improved upon if optical potentials expressed in terms of isoscalar and isovector components such as in Ref. \cite{Bauge01,Jeukenne77} are employed. In such a formulation, we expect the uncertainties in the charge-exchange cross sections to be reduced. The development and implementation of such an optical potential will be the focus of future work.

All interactions considered for this work are purely local, but it will be interesting to include the effects of non-local interactions, since they have been shown to have a large impact on other reaction channels (e.g. \cite{Titus20162}).
The new code {\sc chexnp}, developed during this study, has the capability of incorporating several global optical potentials, including non-local potentials, since it uses the non-local Schr\"{o}dinger equation solver from the NLAT reaction code \cite {Titus2016}. Moreover, our current implementation assumes the reaction takes place in a single step. While the single-step approximation is expected to be valid at medium to high beam energies, this study includes beam energies as low as 25 MeV, where there may be contributions to the charge-exchange cross section from higher-order processes. There have been efforts to quantify effects from multistep processes in (p,n) charge-exchange, including work by Madsen et al. which proposes an energy dependent correction factor \cite{Madsen1980}. Additionally there have been studies of multistep mechanisms for charge-exchange reactions with heavier probes such as ($^3$He,t) \cite{Coker73}, ($^7$Li,$^7$Li) \cite{CLARKE86}, ($^{12}$C,$^{12}$N) \cite{Lenske89}. Future plans for {\sc chexnp} include the extension to multi-step processes.

\begin{acknowledgments}

We are grateful to Amy Lovell, Chlo\"{e} Hebborn, Remco Zegers, and Pawel Danielewicz for many useful discussions and valuable insights. This work was supported in part by the U.S. Department of Energy (Office of Science, Nuclear Physics) under grant DE-SC0021422 and by the National Nuclear Security Administration under the Stewardship Science Academic Alliances program through the U.S. DOE Cooperative Agreement No. DE-FG52-08NA2855. This work was performed under the auspices of the U.S. Department of Energy by Lawrence Livermore National Laboratory under Contract No. DE-AC52-07NA27344.
\end{acknowledgments}

%\bibliography{CHEX_Results}

\providecommand{\noopsort}[1]{}\providecommand{\singleletter}[1]{#1}%

\end{document}